# Forensic Issues and Techniques to Improve Security in SSD with Flex Capacity Feature

Na Young Ahn[1], and Dong Hoon Lee[2], Member, IEEE
[1] Institute of Cyber Security & Privacy, Korea University, Seoul, South Korea
[2] Graduate School of Information Security, Korea University, Seoul, South Korea

Corresponding author: Dong Hoon Lee (e-mail: donghlee@ korea.ac.kr).

The study was funded by Institute for Information and communications Technology promotion (Grant No. 2020-0-00374, Development of Security Primitive for Unmanned Vehicles).

**ABSTRACT** Over-provisioning technology is typically introduced as a means to improve the performance of storage systems, such as databases. The over-provisioning area is both hidden and difficult for normal users to access. This paper focuses on attack models for such hidden areas. Malicious hackers use advanced over-provisioning techniques that vary capacity according to workload, and as such, our focus is on attack models that use variable over-provisioning technology. According to these attack models, it is possible to scan for invalid blocks containing original data or malware code that is hidden in the over-provisioning area. In this paper, we outline the different forensic processes performed for each memory cell type of the over-provisioning area and disclose security enhancement techniques that increase immunity to these attack models. This leads to a discussion of forensic possibilities and countermeasures for SSDs that can change the over-provisioning area. We also present information-hiding attacks and information-exposing attacks on the invalidation area of the SSD. Our research provides a good foundation upon which the performance and security of SSD-based databases can be further improved.

**INDEX TERMS** Forensic, Over-provisioning, Hidden area, Attack model, Malware, NAND flash memory, SSD, Invalid block

## I. INTRODUCTION

In general, a NAND flash-based storage device performs a read/program operation in units of pages and an erase operation in units of blocks [1]. Here, a block is composed of a plurality of pages. Due to a structural issue of NAND flash memory, a data-use unit and a data-erase unit are different from each other. These differences complicate the management of data in NAND flash memory as garbage collection is inevitable for a new block to be created by collecting only valid page data of different blocks [2] and garbage collection naturally extends write amplification [3]. Here, write amplification has a significant impact on the lifetime/endurance of the storage device. In general, garbage collection is performed as a background operation rather than a read/program operation. However, as garbage collection is interrupted by the host's write request, system performance experiences a slowdown.

Data-hiding attacks hide malicious code according to file system manipulation on a flash-based storage device, for example, a USB device [4]. In these types of attacks, the attacker stores malicious code in the master boot record (MBR), the volume boot record (VBR), or another reserved sector that is hidden outside the partition. In addition, the attacker stores this code in the sector for "bad" or "in use" data and manipulates the metadata of the file system, thus preventing the operating system from accessing the stored sector. Recently, forensic techniques to deal with these hidden areas have emerged [5,6].

NAND flash memory inherently has a residual issue of original data because a program operation unit and an erase operation unit are different. For example, a write amplification factor (WAF) greater than 1 means that the data of a valid block has been deleted by the user, but that the invalid block still contains the original data. We can illustrate this by considering data deleted from a smartphone. Indeed, it has been confirmed that, in some cases, the actual smartphone data has still not been deleted for more than six months and discoverable by forensics [7]. This problem is bound to exist for all storage devices based on NAND flash memory. When the storage device is applied to the database and even if personal information is deleted by the user, there is a high possibility that it lingers in the invalid block of the storage device [8,9]. Companies in the database business need to guarantee complete deletion of personal information in these invalidated blocks. It is important to note that it is necessary to ensure complete deletion in both the user area of the storage device as well as in the hidden area.

This paper presents an attack model using the variable over-provisioning domain for the first time. Specifically, we introduce data-hiding attacks and personal information-acquiring attacks. In Section II, we explore variations in the over-provisioning area and corresponding forensic targets. Next, Section III presents the attack models, and Section IV presents the forensic process. Section V details techniques to enhance security against these attacks, and finally, we give an overview of our attack simulation environment. This paper sheds light on the increasing need for balance between performance improvement and security in databases. In the conclusion of this paper, we advocate for future follow-up research in this area to improve the security implementations of SSD-based databases.

## II. VARIABLE OVER-PROVISIONING

Over-provisioning is a function in which some space of a storage device is used as a buffer to improve performance [10].

### A. OVER-PROVISIONING

Over-provisioning, in the simplest of terms, allows the number of physical blocks to be greater than the number of logical blocks. In other words, the SSD controller can see a certain percentage of the physical block, but the operating system or file system is reserved. In general, SSD manufacturers already allocate 7%–25% as over-provisioning area. The user may create more over-provisioning area by simply creating partitions smaller than the available physical space. Although the over-provisioning area is not visible at the operating system level, the SSD controller can still see and use it. The main reason that over-provisioning area is needed is to mitigate the limited life cycle of memory cells. The invisible over-provisioning block is automatically replaced and used when the blocks in the space that are visible to the operating system expire.

Garbage collection uses the idle time of the SSD drive in the background to erase "stale" pages. Erase operations run slower than normal data writes, so on a system where the SSD is constantly under a heavy random write load, garbage collection will run out of "free" pages before even erasing "stale" pages. At this time, the flash translation layer cannot follow the random write in the foreground, and the garbage collection performs an erase operation at exactly the same time a foreground write request is received from the host. Therefore, the over-provisioning area acts as a buffer space to absorb a significant volume of write workload. In short, this buffer space provides enough time for garbage collection to catch up with the workload.

The over-provisioning area is a dedicated buffer space where garbage collection is performed. This space is not disclosed to the user and is controlled by the controller of the NAND flash memory.

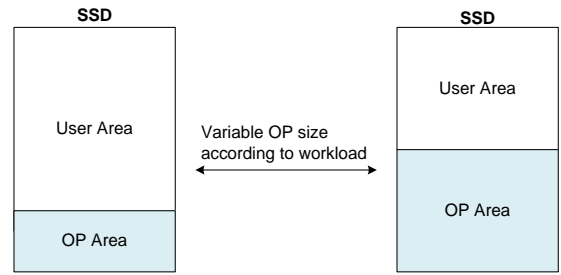

**FIGURE 1.** Variable OP size according to workload.

Referring to Fig. 1, the over-provisioning area is proportional to the performance of the storage device. Recently, researchers have created a technique to vary the over-provisioning area according to the workload in the data storage system [11, 12] by adjusting the physical space to maintain optimal performance depending on the workload. For example, the size of the over-provisioning area depends on the category of the workload [13]. In the case of a workload corresponding to random write requests, the over-provisioning area will be larger than a normal write request. On the other hand, in the case of a workload corresponding to sequential write requests, the over-provisioning area will be smaller than a normal write request. Other researchers have also recently developed a technique to determine a category using machine learning for a channel workload [14, 15], in which the over-provisioning area can be optimized based on the category and expected degradation for the workload.

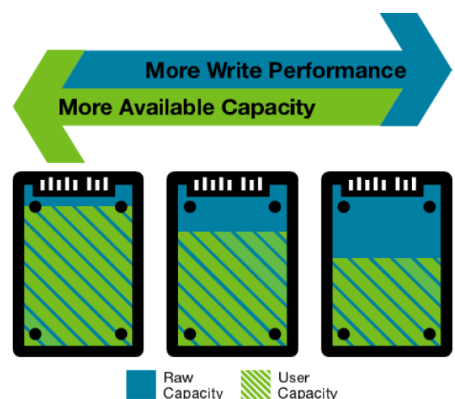

**FIGURE 2.** Flex Capacity Feature in SSD [16].

### B. FLEX CAPACITY FEATURE

Micron recently released an SSD with a flex capacity feature that changes the capacity of the storage device according to the workload [16]. In order to satisfy the needs of various applications, SSDs with flex capacity are expected to demonstrate optimal performance.

Referring to Fig. 2, in this model, there is 1 TB of available space. On the left is an example factory configuration that has 960 GB of usable space. At 960 GB, this SSD offers default Micron user/system capacity, factory-set over-provisioning, and factory-set write IOPS (Input/Output Operations Per Second) performance. The default capacity of 960 GB is a satisfactory default option. The central drive shows how the Micron Flex Capacity feature is used to improve the write performance of an SSD and reduce the user capacity slightly to 800 GB. Write IOPS performance is improved with 800 GB user capacity. Alternatively, if you need to replace an 800 GB SSD, you can easily reset it to 800 GB with a 960 GB Micron SSD.

The drive on the right shows another way to further improve write performance with the Micron Flex Capacity feature by adjusting the user capacity to 480 GB. With this setting, you can now use the same example SSD for more write-intensive workloads or use it to replace a 480 GB SSD that may have been obsolete. In one embodiment, the write IOPS performance is adjusted, or the 960 GB Micron SSD is reset to match the capacity of an existing 800 GB or 480 GB SSD. However, the Micron Flex Capacity feature allows you to easily reset a Micron SSD to 627 GB, 472 GB or any other capacity the user may need.

Although factory and user capacities are different, the same effect can be used for other Micron SSDs with Flex Capacity capability. The Micron Flex Capacity feature allows you to easily change the number of gigabytes available, so you can also choose to perform the write IOPS performance or capacity tuning either permanently or temporarily. SSDs can be set up to best match known workloads, or their characteristics can be altered to manage unexpected application I/O demands more easily. Alternatively, you can increase the write IOPS performance of an SSD either permanently or only when and only for as long as you need. Note that different applications and workloads require different storage for best results. As mainstream data center storage moves rapidly toward SSDs, there is a growing demand for configurations that are optimized specifically for both IOPS performance and user capacity.

With the Micron Flex Capacity feature, planners, architects, implementers, and administrators no longer have to compromise the limited number of SSD configurations, performance options, and capacities. The Micron Flex Capacity feature makes it easy to create application-tunable SSDs from flash. Whether applications and workloads require greater capacity with cost-intensive and read-intensive workloads, greater write IOPS performance for write-intensive workloads, or performance that is suitable for mixed use (including read/write balancing), Flex's capacity capabilities enable precise performance and capacity tuning.

### C. FORENSIC TARGET

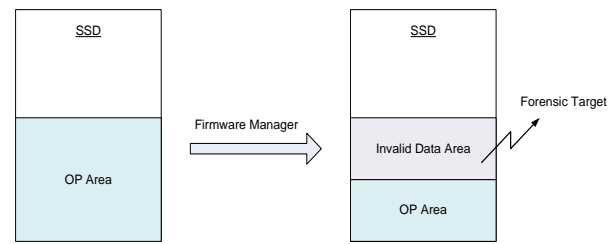

**FIGURE 3.** Forensic Target in OP area.

The firmware manager can adjust the size of the SSD's OP (Over-Provisioning) area. The forensic target is an invalid data area according to the variation of the OP area, as per Fig. 3. This target is considered under the assumption that data in the changed area is not completely deleted even if the OP area is changed. Typically, when changing an OP area, the probability of adding a wiping operation to the changed region is very low.

### III. ATTACK MODEL
### A. INFORMATION-DISCLOSING ATTACK MODEL

Recently, many storage devices vary the size of the OP area in real time to optimize performance. In general, it is known that the larger the size of the OP area, the better the performance. There is a case where the OP area is set by a maximum of 50%. The region we are interested in as a forensic target is the invalidation data region created by varying the OP area. The OP area can be freely changed by the user or by the firmware manager. However, a hacker can also change the size of the OP area to a smaller size by using the firmware manager, as shown in Fig. 4.

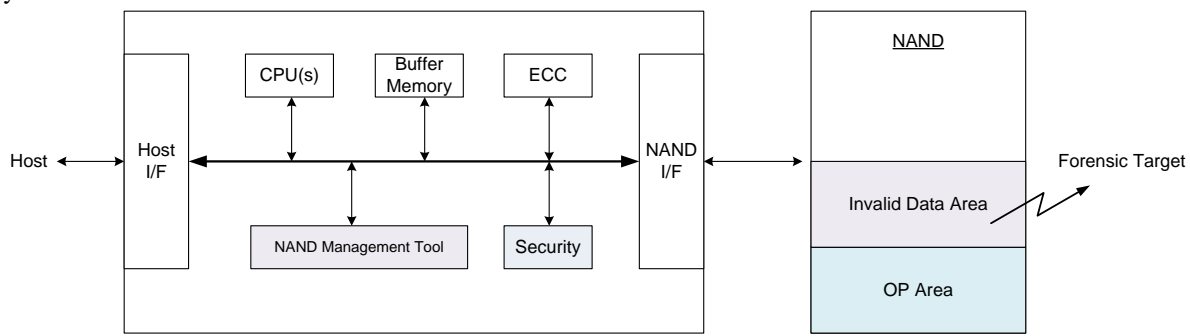

**FIGURE 4.** Information Disclosure Attack Model using variable OP area.

In this process, an invalid data area is inevitably generated. From the data protection point of view, this is a good result because an erase operation on the newly generated invalid data area can be immediately performed. However, according to the management policy of the storage device, the invalid data area is likely to be left unattended for a considerable period. This is because, in terms of management cost, many view it as beneficial to leave the invalid data area as it is rather than to perform an erase operation on a significant invalid data area.

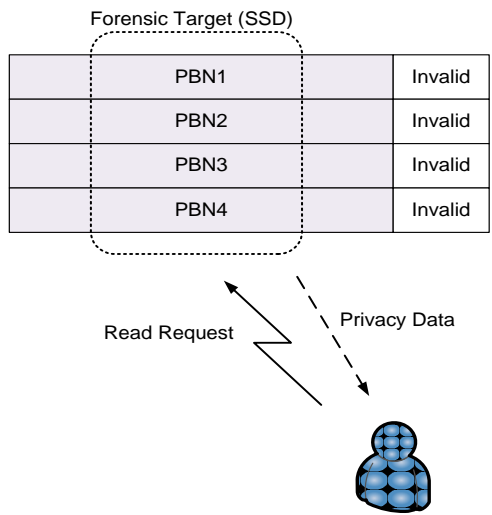

**FIGURE 5.** Forensic Issues on Privacy Data in SSD.

Assuming that the hacker can access the management table of the storage device, the hacker can access this invalid data area without any restrictions. In the related part, data can be secured by accessing the invalid data area of NAND flash memory using forensic equipment. However, in the case of a storage device with a variable OP area, a legitimate read operation can be requested by a host user if the address of the invalid data area is known without special forensic equipment. Without the need for special forensic equipment, as a computer user, a hacker can access these invalid data areas of the NAND flash memory, as shown in Fig. 5. Depending on sensitive information is stored in the invalid data area, computer users can feel more or less alarmed by this.

Such an information-disclosing attack can be forcibly performed at a specific point in time by the firmware and SSD management rights acquirer. If the database implemented by SSD stores sensitive information such as financial information or personal information, the above-described information-disclosing attack would create a serious problem. What's more, the severity of the access to such sensitive information is greater in that it would be untraceable. The acquirer of the firmware and SSD management authority for an OP variable may primarily change the OP area to read sensitive information and secondarily change the OP area after acquiring said sensitive information. The shorter the time required to acquire sensitive information, the shorter the information-disclosing attack time for the OP area. This means that legitimate users or DB managers may not even have enough time to become aware of these attacks.

Therefore, the management of the invalid data area needs to be more thorough. It is necessary to manage not only the invalidation data of the existing user area, but also the invalidation data of the hidden area. This is an unavoidable problem because the change in authority for the hidden area must be opened to improve performance. In a situation in which the hidden area can become the user area at any time, and vice versa, simply breaking the link of the mapping table does not guarantee prevention of original data leakage. Since invalid data includes original data by default, an efficient and complete deletion technique is necessary to prevent original data from being leaked.

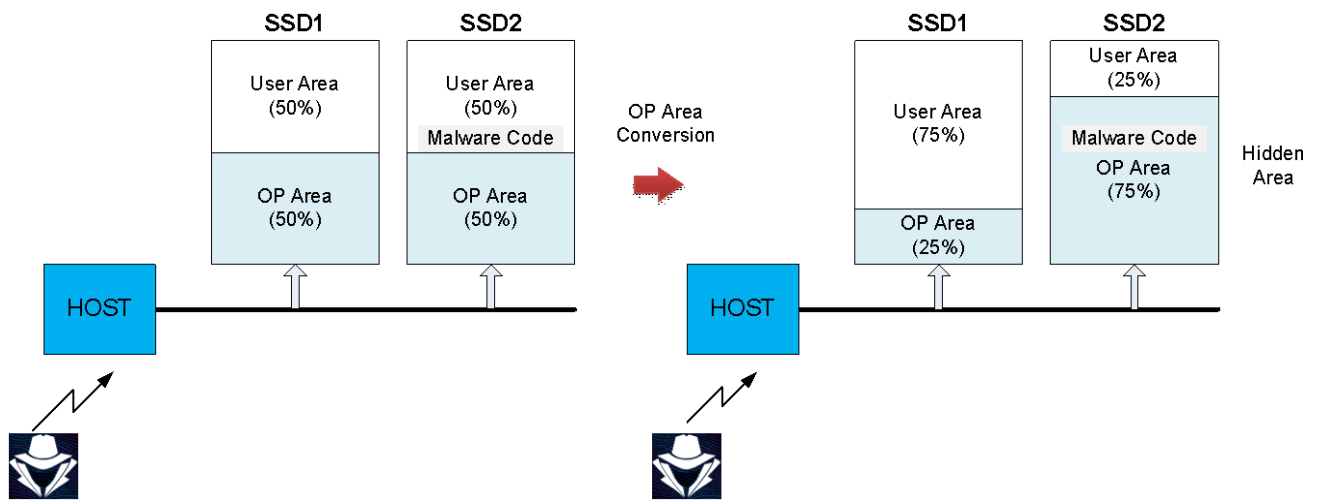

**FIGURE 6.** Malware code injection process using conversion of OP area. SSD1 and SSD2 are connected to one channel. According to this attack, malware code is inserted into the hidden area, i.e. OP area.

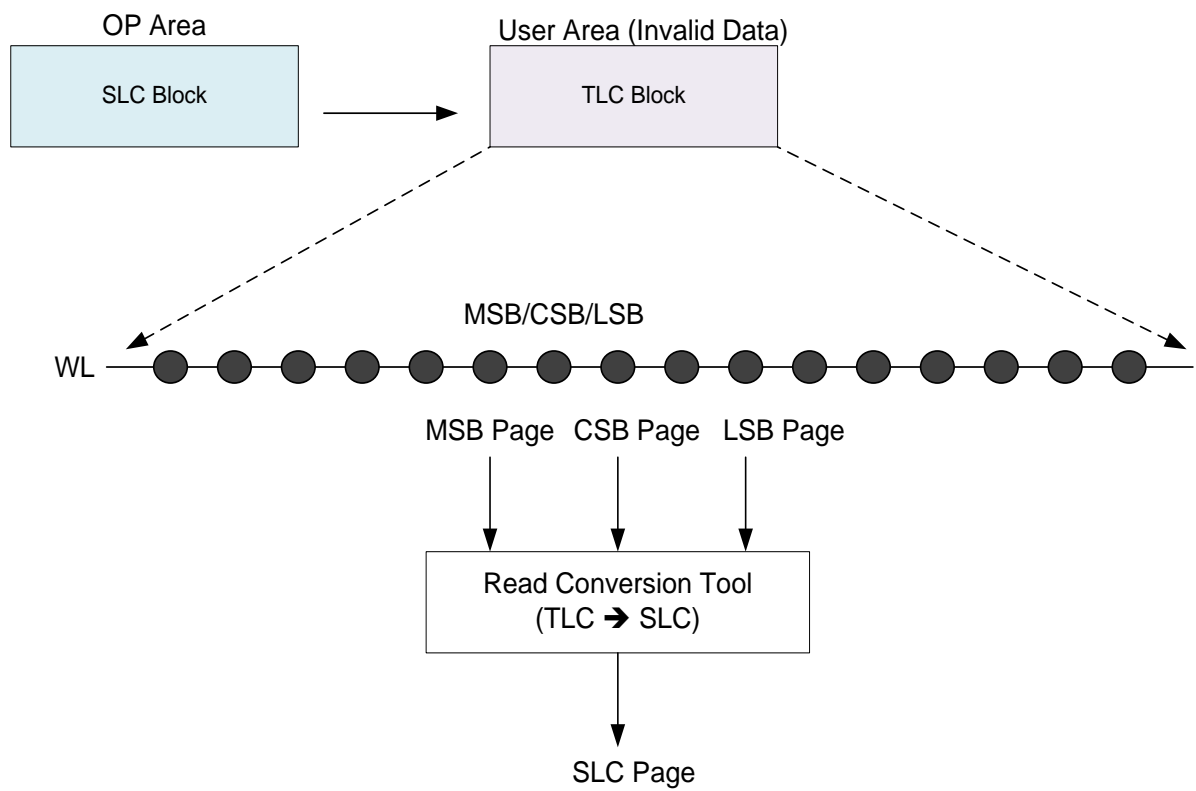

**FIGURE 7.** Forensic Process using Read Conversion Tool. Forensic data is acquired by converting multi-page into single-page.

### B. TEMPERING ATTACK MODEL
Changing the OP area implies partially transferring access rights to the hidden area of the storage device to the user. By using the firmware that manages the size of the OP area, the user can perform any desired operation on the hidden area. The OP area is likely to be used as a user's "secret safe," so to speak, to hide secret information. Assuming that the user has authority over both the firmware and the flash conversion layer, the user can subsequently invalidate the stored secret information after storing it in the user area. According to this invalidation processing operation, the secret information is not physically deleted from the user area and only the mapping table entry becomes deleted. Thereafter, the user changes the area in which the invalidated secret information is stored to the OP area via the firmware. The OP area is a well-known hidden region, and the user's secret information is hidden there.

What would happen if the user were a hacker? As shown in Fig. 6., a hacker may hide malicious code, that is, a malware code, in the OP area. In the drawings, it is assumed that two storage devices SSD1 and SSD2 are connected to a channel in order to simplify the description. Each storage device has 50% OP area. After the hacker stores the malware code in SSD2, they immediately reduce the OP area of SSD1 to 25% and expand the OP area of SSD2 to 75%. At this time, the malware code is included in the hidden area of SSD2. A hacker who gains access to the SSD can activate the embedded malware code at any time by resizing the OP area. Since normal users maintain 100% user area on the channel, it will not be easy to detect such malicious behavior of hackers. Therefore, research to detect or prevent malicious concealment of such hackers should be continued.

The severity of this tempering attack is that it is unlikely that a normal user or a legitimate DB administrator will recognize the attack. Once the malicious code is inserted into the hidden area, such attacks cannot be detected because normal users cannot access the hidden area. In addition, even for a legitimate DB administrator, the possibility of detection is very low because the malicious code inserted in the hidden area is stored in the form of invalid data rather than valid data. Therefore, considering the severity of such a tempering attack, there is a growing need for techniques that can detect malicious codes not only in the user area but also in the hidden area as well as in the valid data area and the invalid data area.

No malicious code detection technique for the invalid data area is provided by the standard operation of existing NAND flash memory. Although detection is possible through various forensic equipment and techniques, it would be incredibly time-consuming and costly for DB operators to apply these in real time. Accordingly, a NAND flash memory provider needs to change the standard protocol to perform a scan/read operation on an invalid data area. Here, the scan read operation includes a read operation on the invalid data area according to a special request of the host.

In order for the user to make such the special request, the SSD needs to monitor the ratio of the valid data area to the invalid data area. When the ratio of the invalid data area to the valid data area exceeds a certain reference value, the SSD may warn the host about the special request. The host may also receive a warning when there is a sharp change in these ratios. Since it is difficult to detect the tempering attack described above, the ratio of the invalidation data area and the validated data area must be monitored in real time. Future enterprise SSDs will need to have a special processor to prevent such tempering attacks.

## IV. FORENSIC PROCESS

The memory cells comprising blocks in the OP area may be different according to NAND manufacturers. In general, memory cells constituting a NAND flash memory include SLC(Single Level Cell), MLC(Multi Level Cell), TLC(Triple Level Cell), QLC(Quad Level Cell), PLC(Penta Level Cell) and the like [1-3]. SLCs store one data bit in one memory cell, MLCs store two bits in one memory cell, TLCs store three data bits in one memory cell, and QLCs store four data bits in a cell. The more bits stored in one cell, the more time it takes to write or read data, and the more difficult it is to manage. There are products that manage the OP area using SLC, and there are products that manage the OP area with the corresponding MLC/TLC/QLC cell type to the user area memory block. Therefore, the forensic process is different depending on the type of memory cells constituting the OP area.

### A. SLC OP AREA

In the case of SLC, when the OP area is changed to the user area, the capacity of the invalid block will increase three-fold. A hacker can perform a read data for about three pages (MSB/CSB/LSB) to one physical page. The three-page read data may be changed to one page to correspond with the SLC program. As shown in Fig. 7, this page data is composed of a combination of data in an erased state and data in the remaining states.

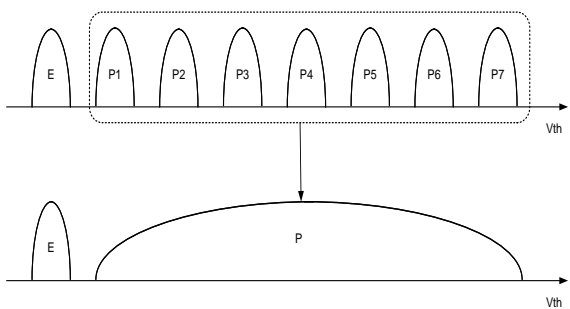

**FIGURE 8.** State Mapping when changing 3-pages to 1-page.

The read conversion tool may receive MSB(Most Significant Bits) page data, CSB(Center Significant Bits) page data, and LSB(Least Significant Bits) page data, and it may output one SLC page data. Note that the conversion principle is simply to replace the erase state (E) and the remaining states (P1 to P7) among the eight states (E, P1 to P7) of the TLC with two SLC states (E, P) as shown in Fig. 8.

### B. TLC OP AREA

When the OP area is configured as MLC/TLC/QLC, the hacker will perform a normal read operation on the invalidated block changed from the OP area to the user area. At this time, it is assumed that the hacker knows the exact physical address of the changed invalid block. The hacker may transmit a read request for the target area to the SSD. In this case, the read request may be a specific file name, and the extension may be 'txt'. For example, the hacker makes a read request for 'dump.txt' in the target area. In response to a read request for 'dump.txt', the SSD may perform a read data for the target area of the invalid data area and transmit the read data to the hacker.

## V. PROPOSED SECURITY ENHANCEMENT SCHEMES

As above mentioned, existing variable OP technologies have forensic issues. Even without advanced forensic skills, the possibility of forensics via a hacking tool by a user is significantly high. For storage systems used for databases, this can be a serious threat. Against these threat agents, we suggest the following security enhancement techniques. We believe that these techniques should be mandatory and not optional.

### A. DATA ENCRYPTION & STRONG AUTHENTICATION

FIPS140-2 is the US government standard that sets out encryption and related security requirements with which IT products must demonstrate compliance when used for non-confidential and sensitive purposes. Generally speaking, SSDs with encryption built into their hardware are more commonly referred to as self-encryption drives (SED). SED technology delivers proven and certified data security that provides virtually impenetrable pre-boot access protection to user data [17]. Because encryption is part of the drive's controller, it provides pre-boot data protection. Attempting to run a software utility to crack the verification code is impossible because encryption is enabled before the software is loaded. Another advantage of always-enabled encryption is that it complies with the TCG(Trusted Computing Group) Opal 2.0 specification and IEEE-1667 access authentication protocol to ensure that the drive meets government compliance requirements for data in banking, financial, medical, and government applications [17, 18]. However, a problem has been revealed that connects to the JTAG debugging interface of the drive and modifies the password verification routine to always maintain successful authentication regardless of the entered password.

Security functions in the existing SSD largely include the above-described SED and TCG. These security functions are simply techniques that secure data confidentiality. Many SSD manufacturers and database customers remain silent on the previously mentioned forensic issues for invalidated data areas. But this silence cannot be the answer. This is because serious information disclosure or information hiding can be achieved rather easily as the OP area is varied. For example, even if all processed data is encrypted, a hacker can always hide data using the invalidated area. Furthermore, even if it is encrypted data, a hacker with quantum computing technology can secure encrypted data via a read operation for the invalidation area and perform a decryption operation on the secured data. In order to be free from such forensic issues, deletion verification for the invalidated area should be mandatory as an essential process. Therefore, in order to fundamentally block the above-mentioned problems, access to the firmware that varies the OP area should be more strongly regulated or stronger methods for authenticating SSD firmware users should be required.

### B. SECURE GARBAGE COLLECTION

Basically, NAND flash memory causes a significant amount of garbage collection because the program/read unit and the erase unit are different. Such garbage collection inevitably follows the propagation of the original data inside NAND flash memory [19]. In other words, if the original data exists in the validated block, there is a very high possibility that the original data is left as is in the plurality of invalidated blocks. From a privacy point of view, garbage collection should be performed to prevent the spread of original data. Ahn was the first to propose this discussion and introduced a secure garbage collection method that prevents the spread of original data via a novel secure copy-back program [8, 9, 19].

### C. PSEUDO ERASE OPERATION

In order to improve the real-time performance of database systems, flexible OP functions are also essential. In a system equipped with such functions, the erase operation is required in an on-the-fly manner for blocks changed from the OP area to the user area. A normal erase operation is not recommended due to cost and time. Accordingly, it is important to develop a pseudo erase operation that reduces time and cost burdens. A pseudo-erase operation should, at the base, include a full or partial program operation [8, 9, 20, 21, 22]. The large-capacity pseudo erase operation may be a sufficient issue as a future research project.

### D. ZONE NAMESPACE

In general, when garbage collection is performed, additional write operations occur. Garbage collection increases Write Amplification Factor (WAF) by default. Although a general storage device stores data in target areas logically corresponding to a plurality of application programs, in reality, data is sequentially stored in one block of a NAND flash memory. Since data is stored regardless of its properties, there is a high possibility of garbage collection occurring according to data changes. In order to reduce the possibility of garbage collection, the Zone Namespace (ZNS) solution performs a function that allocates storage area by data properties [23, 24]. In general, ZNS solution has the following advantages: First, the lifespan of the storage device is increased. Second, it improves write speed in a multi-user environment because the likelihood of garbage collection is reduced. Third, the over-provisioning area is reduced. Fourth, the DRAM can be used efficiently because the mapping table management usage is reduced.

### E. VERIFICATION OF DATA DELECTION AND RELATED NEW COMMAND

The SSD forensic issue is stems from its inability to completely erase data. Although the NAND flash erasure operation exists, only the effective data area becomes erased. Even in the invalid data area, it is difficult with existing technologies to ensure that the original data is deleted. Accordingly, there is a need to improve the security of NAND flash memory so that the secure deletion of original data can be guaranteed. This deletion verification operation should be further discussed and studied more in depth. In addition, there is a growing demand for a new command in relation to deletion verification.

### F. VALID/INVLID DATA RATE MONITORING

As mentioned earlier, tempering attacks and data-disclosing attacks are difficult to detect. In addition, the time and cost required to discover these attacks is significant. Therefore, in order to provide an SSD that is safe from such attacks, it is necessary to monitor the valid/invalid data ratio inside the SSD in real time. When there is a sudden change in the ratio, or when the invalid data ratio increases significantly, the SSD determines that the threat of a tempering attack or a data-disclosing attack is significant.

### G. MANAGER'S INTEGRITY

It is not easy for a hacker to gain administrator privileges. However, against these untraceable attacks by an internal administrator, there is a possibility that the above-mentioned attacks are executed with malicious intent. For one, the SSD-based database administrator or SSD maintenance staff need to secure royalties. Even if you are not a malicious hacker, a misguided employee can easily free hidden information and leak it by using the OP area variable firmware/software at any time. Indeed, this scenario is always the most difficult and challenging topic to mitigate in security studies.

### H. AI-BASED ATTACK DETECTION

It should not be overlooked that these attacks could very well be lodged by artificial intelligence (AI) in lieu of humans. AI can perform the above-mentioned attacks on the SSD invalidation area very carefully, in an instant, and at any moment. This fact notifies that data packets transmitted and received on the SSD channel and monitoring of firmware settings should be performed in real time. In the future, we anticipate that a special-purpose processor for SSD security will be needed. Such a processor will be implemented to detect attacks on the invalidated area of the SSD based on AI and to respond quickly to such security events.

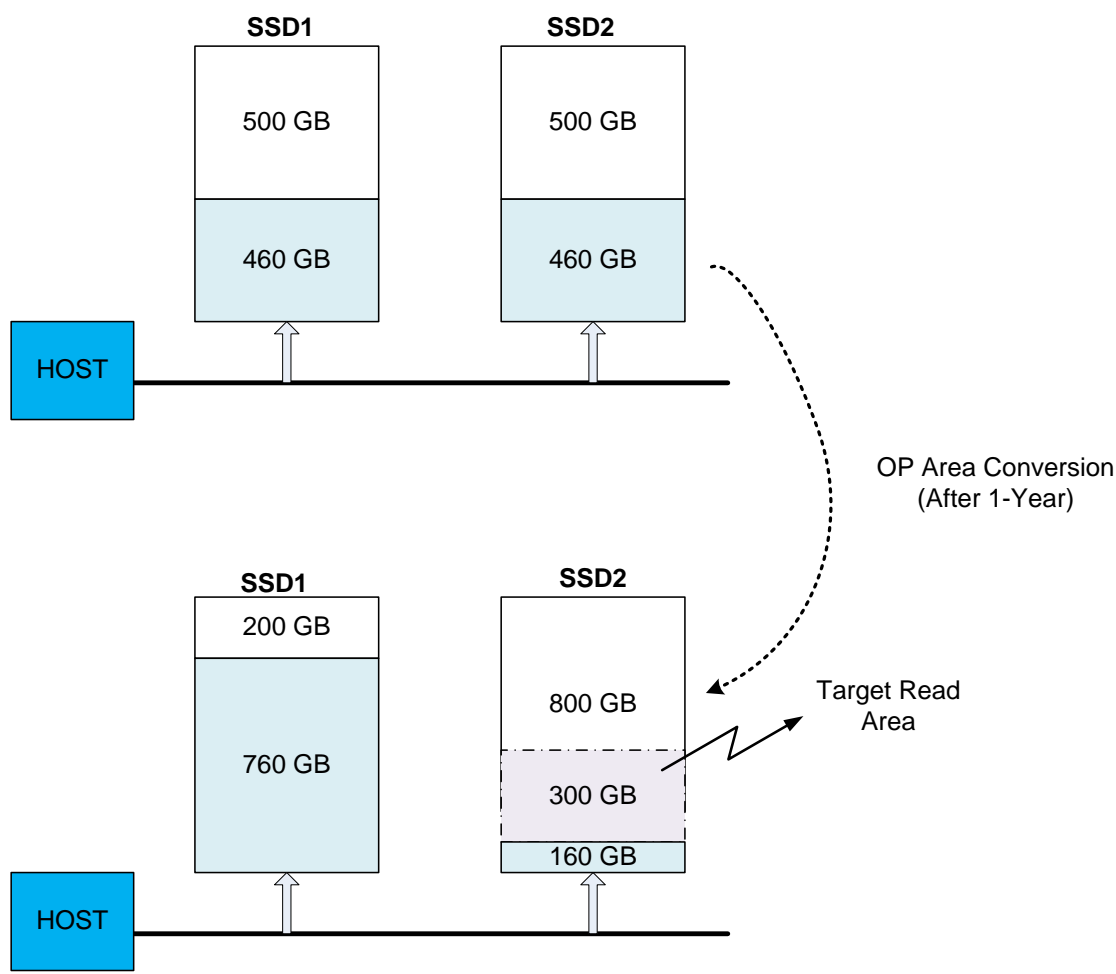

FIGURE 9. Database-like Attack Simulation Environment. 960GB SSDs are connected to one channel. After one year, the OP area is largely changed, and a scan read operation is performed on the changed invalidation area.

## VI. ATTACK SIMULATION ENVIRONMENT

The attack simulation environment consists of two storage devices (SSD1, SSD2) connected to the host device. The host device may vary the OP area of the storage devices SSD1 and SSD2 by firmware. For convenience of explanation, the host device is set to use 1,000 GB of user area. For storage, we want to use 960 GB of Micron 5200 series products. The host device is configured to be suitable for an environment that provides an image-based SNS service.

Before the membership-based image-sharing social media service is provided, the OP area is set for both SSD1 and SSD2 with the same capacity. For example, SSD1 and SSD2 are both set to an OP area of 460 GB, referring to Fig. 9. Thereafter, one year after the start of providing the SNS service, the OP area is changed by the firmware. For example, SSD1 is set to have an OP area of 760 GB, and SSD2 is set to have an OP area of 160 GB. Thereafter, a scan read operation is performed on the 300 GB area newly incorporated into the user area in SSD2. Analyze the data according to the scan read operation.

In addition to this, it is necessary to verify the experiment when the storage device is applied in the member DB. In the case of member DB, there is a greater possibility of privacy exposure, so it is necessary to proceed in future research.

The simulation environment described above is significantly different from the actual data center environment. However, since Micron 5200 series products with flex capacity features that are actually applied to data centers. As such, the analysis of threats to information-disclosing attacks and tempering attacks and the subsequent results can be considered as sufficiently tested. Similar experiments can be added not only with Micron SSD products but also with products from other NAND vendors. In addition, the above-described simulation may be performed in an environment using SSD products from different companies. These various simulations will be of sufficient value for future research.

In addition, there is a need to proceed with research to create a simulation test environment. Monitoring an SSD for one year is impossible on a practical level. Technology that is capable of accelerating input/output for data of a specific pattern could be applied to perform a simulation. For example, the above-described test could be performed by making an image file of an SSD that has passed a certain time in an actual product and applying this image file.

## VII. CONCLUSION

We presented attacks using the variable OP area of SSD. For example, we discussed an information-disclosing attack model and a tempering attack model—both of which vary the OP area. The information-disclosing attack model was implemented as a forensics technology for the invalid data area created by changing the OP area. The tempering attack model consisted of inserting malicious code into a hidden area such as the OP area. In addition, a forensic process is presented according to the type of memory cell in the OP area. In the case of the SLC OP area, the forensic process converted a three-page read data for the invalid data area into a single-page read data. In the case of the TLC OP area, the forensic process executes as a normal read operation for the invalidation data region. In addition, we presented various security enhancement techniques for improving security against attacks utilizing the variable OP area.

Finally, we proposed an attack simulation environment similar to the DB system environment along with a simulation method. SSD-based databases vary the OP area for performance purposes, and according to this variable operation, this paper confirms that the SSD is vulnerable to information-hiding attacks and information-exposing attacks on the invalidated data area. While the two attack models we presented are only examples, increasingly diverse attacks on the invalidation area are emerging. We presented several new attack techniques and countermeasures to improve the security of the existing SSD, and we believe that these need to be applied immediately in the field. In the future, there will be a need for more detailed and advanced follow-up studies on the invalidation area of these SSDs. Our research will be an essential addition to the DB system, to which deletion obligation is to be applied. We also predict that additional studies on detecting attacks on the OP area and verifying original data deletion will be demanded.

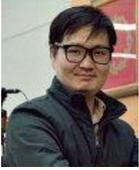

**Na Young Ahn** is a post-doc researcher with the Institute of Cyber Security & Privacy at Korea University, South Korea. He holds a Ph.D. in Cyber Security. He received his B.S. and M.S. degrees from the Department of Electrical Engineering at Korea University. He has been a patent engineer at patent and law firms since 2005. His articles have been published in journals including IEEE Access and Ad hoc & Sensor Wireless Networks. His research interests include physical layer security in vehicular communications, biometric authentications, Non-Competitive blockchain, and anti-forensics in flash memories.

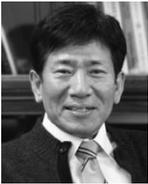

**Dong Hoon Lee** received his B.S. degree in economics from Korea University, Seoul, Korea, in 1985 and M.S. and Ph.D. degrees in computer science from the University of Oklahoma, Norman, OK, USA, in 1988 and 1992, respectively. Since 1993, he has been with the Faculty of Computer Science and Information Security, Korea University. His research interests include the design and analysis of cryptographic protocols in key agreement, encryption, signatures, embedded device security, and privacy-enhancing technology.